\documentclass[letter,twocolumn]{jpsj3}
\usepackage{txfonts}
\usepackage{color}
\title{Crystalline Electric Field and Kondo Effect in SmOs$_4$Sb$_{12}$}
\author{Shota Mombetsu$^1$\thanks{mombetsu1412@gmail.com}, Tatsuya Yanagisawa$^1$, Hiroyuki Hidaka$^1$, Hiroshi Amitsuka$^1$, \\
Shadi Yasin$^{2,3}$, Sergei Zherlitsyn$^3$, Jochen Wosnitza$^3$, \\
Pei-Chun Ho$^4$, and M. Brian Maple$^5$}
\inst{$^1$Department of Physics, Hokkaido University, Sapporo 060-0810, Japan \\
$^2$American University of the Middle East, College of Engineering and Technology, Eqaila, Kuwait \\
$^3$Hochfeld-Magnetlabor Dresden (HLD-EMFL), Helmholtz-Zentrum Dresden-Rossendorf (HZDR), D-01314 Dresden, Germany \\
$^4$Department of Physics, California State University, Fresno, CA 93740, U.S.A. \\
$^5$Department of Physics and Center for Advanced Nanoscience, University of California San Diego, La Jolla, CA 92093, U.S.A.} 

\abst{Our ultrasound results obtained in pulsed magnetic fields show that the filled-skutterudite compound  SmOs$_4$Sb$_{12}$ has the $\Gamma_{67}$ quartet crystalline-electric-field ground state. This fact suggests that the multipolar degrees of freedom of the $\Gamma_{67}$ quartet play an important role in the unusual physical properties of this material. On the other hand, the elastic response below $\approx$ 20 T cannot be explained using the localized 4$f$-electron model, which does not take into account the Kondo effect or ferromagnetic ordering.
The analysis result suggests the presence of a Kondo-like screened state at low magnetic fields and its suppression at high magnetic fields above 20 T even at low temperatures.}

\begin{document}
\maketitle
\def\vector#1{\mbox{\boldmath $#1$}}
The filled-skutterudite compound SmOs$_4$Sb$_{12}$ has attracted much attention because of its unusual physical properties \cite{Sanada,Yuhasz}. Namely, this compound shows a relatively large Sommerfeld coefficient, $\gamma$ $\approx$ 820 -- 880 mJ mol$^{-1}$ K$^{-2}$, which is hardly affected by magnetic fields of up to 8 T \cite{Sanada}. This robust $\gamma$ is in contrast to the magnetically suppressed $\gamma$ in typical Ce-based heavy-fermion compounds such as CeCu$_6$ \cite{CeCu6}. This suggests that the heavy-fermion state in SmOs$_4$Sb$_{12}$ is of non-magnetic origin. On the other hand, several phenomena due to the anharmonic vibrations of guest ions, so called rattling, have been observed in this material \cite{Raman,INS,Einstein,SmOs4Sb12_Yana}. A multichannel Kondo effect due to the coupling between the rattling modes and conduction electrons has been theoretically proposed to explain the magnetically robust heavy-fermion state \cite{4Level,Inv}. The characteristic temperature $T^\ast$, which is related to the crossover temperature of Kondo-like screening due to hybridization, is estimated to be $\approx$ 20 K \cite{NQR_R_JPSJ}. Below $T^\ast$, SmOs$_4$Sb$_{12}$ shows a phase transition at $T_{\textrm C}$ $\approx$ 2.6 K, accompanied by the appearance of a weak spontaneous magnetic moment of 0.02 -- 0.03 $\mu_{\rm B}$/Sm-ion \cite{Sanada,Yuhasz}, which is much smaller than the free-ion value of 0.71 $\mu_{\rm B}$/ion for Sm$^{3+}$, even considering an intermediate valence of $\approx$ +2.76, as found by X-ray absorption spectroscopy below 20 K \cite{Valence}. One of the possible explanations is that the low-temperature phase is an itinerant ferromagnetic state \cite{Sanada}. Another possible explanation would be a contribution of higher-order multipole moments to the low-temperature properties, which has been suggested on the basis of a hydrostatic-pressure study \cite{UnderP_C_AcChi}. 
\begin{figure}[t]
\begin{center}
\includegraphics[width=0.7\linewidth]{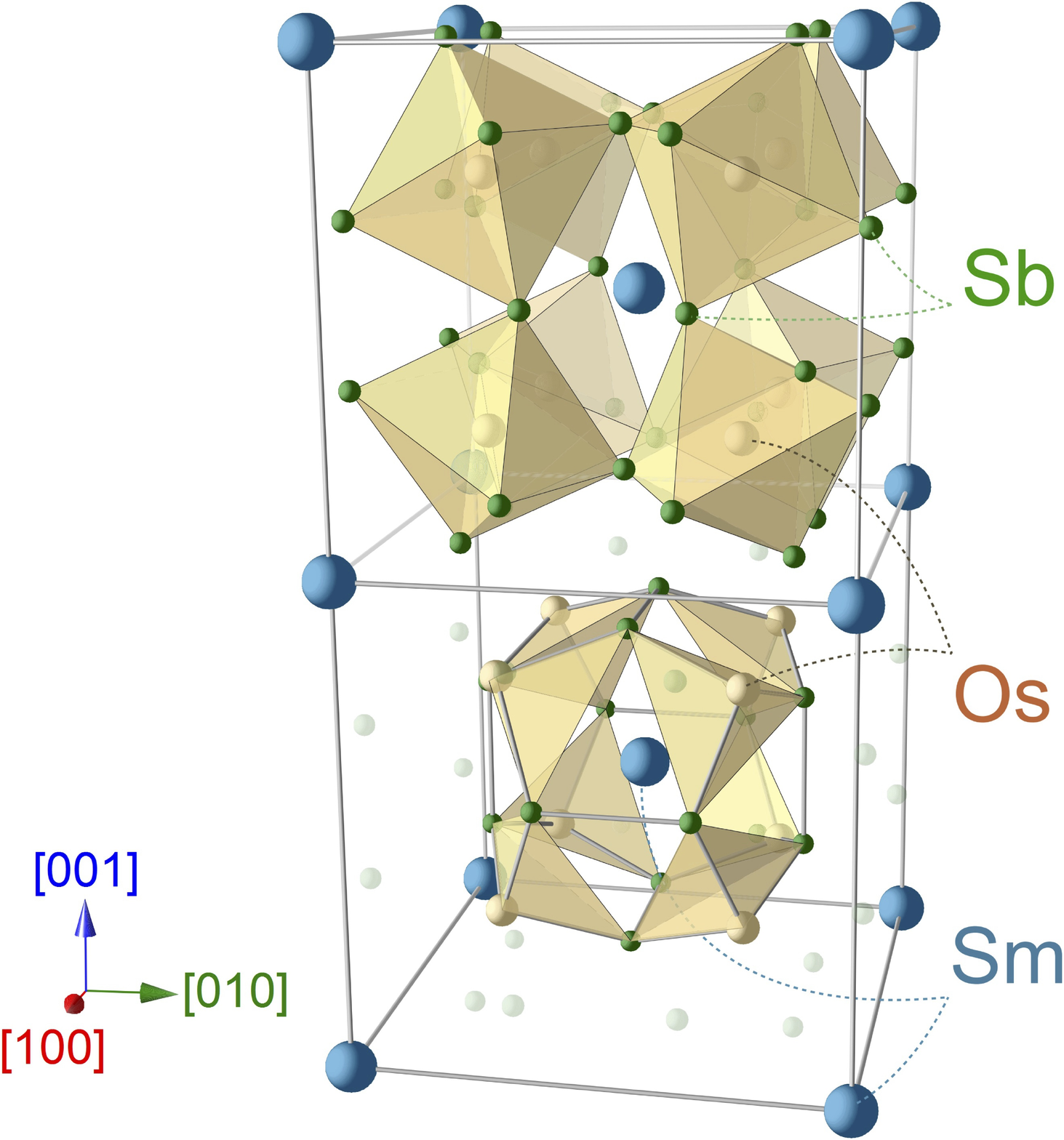}
\end{center}
\caption{(Color online)  Crystal structure of SmOs$_4$Sb$_{12}$ (space group: Im$\bar{3}$), where two unit cells are described. In the upper cell, an octahedra that consists of Sb atoms around an Os atom is shown. In the lower cell, a cage structure that consists of Os and Sb atoms is emphasized.}
\label{f1}
\vspace{-5mm}
\end{figure}

\begin{table}[t]
\caption{Quadrupoles, symmetrized strains, and corresponding elastic constants are categorized with $T_{\textrm h}$ symmetry ($O_{\textrm h}$ symmetry).}
\label{t1}
\begin{center}
\begin{tabular}{llll}
\hline
\multicolumn{1}{c}{Symmetry} & \multicolumn{1}{c}{Quadrupole} & \multicolumn{1}{c}{Strain} & \multicolumn{1}{c}{Elastic Constant} \\
\hline
$\Gamma_1$ &  & $\varepsilon_B = \cfrac{\varepsilon_{xx}+\varepsilon_{yy}+\varepsilon_{zz}}{\sqrt{3}}$ & $C_B = \cfrac{C_{11}+2C_{12}}{3}$ \\
$\Gamma_{23}$ $(\Gamma_{3})$ & $O^{0}_{2}, O^{2}_{2}$ & $\cfrac{2\varepsilon_{zz} - \varepsilon_{xx}-\varepsilon_{yy}}{\sqrt{6}}, \cfrac{\varepsilon_{yy}-\varepsilon_{xx}}{\sqrt{2}}$ & $\cfrac{C_{11}-C_{12}}{2}$ \\
$\Gamma_{4}$ $(\Gamma_{5})$ & $O_{yz}, O_{zx}, O_{xy}$ & $\varepsilon_{yz}, \varepsilon_{zx}, \varepsilon_{xy}$ & $C_{44}$ \\
\hline
\vspace{-3mm}
\end{tabular}
\end{center}
\end{table}

In some Sm-based cage-structured compounds with a cubic symmetry and a quartet ground state, such as SmRu$_4$P$_{12}$ and SmTi$_2$Al$_{20}$, the possible effects of multipolar degrees of freedom have been discussed \cite{Octupole_SmRu4P12, SmTi2Al20}. In several Pr-based filled-skutterudite compounds with a pseudodegenerate crystalline-electric-field (CEF) ground state, such as PrRu$_4$P$_{12}$, multipole ordering has been suggested \cite{PrRu4P12}. Because of the similar cage-featured crystal structure (Fig. 1), it is expected that the unusual physical properties of SmOs$_4$Sb$_{12}$ also originate from higher-order multipoles.

To study the role of multipolar degrees of freedom in this material, knowledge of the CEF level scheme of the Sm$^{3+}$ ion is required.
Owing to CEFs in the cubic $T_{\textrm h}$ symmetry, the 6\--fold degenerate ground-state multiplet $\vector{J} = 5/2$ of the Sm$^{3+}$ ion splits into a $\Gamma_{67}$ quartet and a $\Gamma_{5}$ doublet. The $\Gamma_{67}$ quartet has not only magnetic dipoles but also electric quadrupoles and magnetic octupoles, while the $\Gamma_{5}$ doublet has only magnetic dipolar degrees of freedom \cite{Shiina}. 
In this study, we have investigated the CEF ground state of SmOs$_4$Sb$_{12}$ by ultrasonic measurement.
There are three independent components of elastic strains in the $T_{\textrm h}$ ($O_{\textrm h}$) symmetry, as listed in Table I, and one can obtain quadrupolar susceptibilities from elastic-constant measurements \cite{QS}. Since quadrupolar susceptibilities strongly depend on the CEF splitting of the 4$f$ states, one can discuss the CEF level scheme on the basis of an ultrasonic study.

\begin{figure}[t]
\begin{center}
\includegraphics[width=1.0\linewidth]{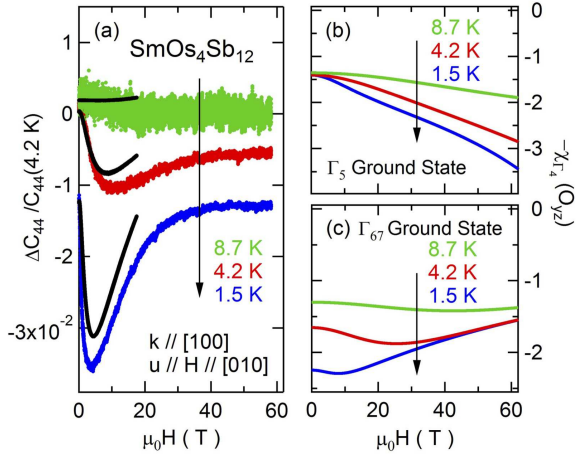}
\end{center}
\caption{(Color online) (a) Relative change in the elastic constant, ${\Delta}C_{44}/C_{44}$, versus magnetic field. The pulsed field data ( $\vector{k}$ $\parallel$ [100] and $\vector{u}$ $\parallel$ $\vector{H}$ $\parallel$ [010], color plots) and static field data ($\vector{H}$ $\parallel$ $\vector{k}$ $\parallel$ [001] and $\vector{u}$ $\parallel$ [010], black curves) are shown.
The zero-field values of the experimental results are calculated on the basis of the temperature dependence of $C_{44}$, which is shown in Fig. 4(a).
(b) Magnetic field dependence of the quadrupolar susceptibility ${\chi}_{{\Gamma}_4}$($O_{yz}$), calculated for the $\Gamma_{5}$ ground state and the $\Gamma_{67}$ first excited state, separated by a gap ${\Delta}$ ${\approx}$ 38 K. (c) The same as in (b) but for the $\Gamma_{67}$ ground state and the $\Gamma_{5}$ first excited state separated by ${\Delta}$ ${\approx}$ 20 K (see text for details).
}
\label{f1}
\end{figure}

Thus far, two models of the CEF level scheme have been proposed for SmOs$_4$Sb$_{12}$. A $\Gamma_{5}$(0 K)$-$$\Gamma_{67}$(38 K) model was suggested from specific heat and electrical resistivity measurements \cite{Yuhasz}. On the other hand, a $\Gamma_{67}$(0 K)$-$$\Gamma_{5}$(20 K) model was suggested on the basis of a point-charge model using extrapolated parameters of PrOs$_4$Sb$_{12}$ \cite{Sanada}. Furthermore, anisotropic features in the magnetization provide evidence of a $\Gamma_{67}$ ground state \cite{Mag_Anisotropy}. However, the presence of Kondo-like screening below $T^\ast$ $\approx$ 20 K and the ferromagnetism below $T_{\textrm C}$ $\approx$ 2.6 K makes it difficult to determine which of the CEF energy level schemes is correct by performing measurements at low temperatures and low magnetic fields. Note also that inelastic neutron scattering is challenging for this compound, since the naturally abundant Sm atom has a large neutron absorption cross section.

Ultrasonic measurements have successfully elucidated low-lying CEF states in several Kondo systems \cite{US_CeB6, US_PrOs4Sb12}.
In a previous report on the ultrasonic study of SmOs$_4$Sb$_{12}$, the CEF state has been discussed, but the determination is difficult owing to the possible presence of the strong screening of quadrupole moments below $T^\ast$ \cite{SmOs4Sb12_Yana}.
In this study, we focus on the quadrupolar susceptibility of SmOs$_4$Sb$_{12}$ at high magnetic fields to determine the CEF states without the effect of the Kondo-like screening.
As will be shown later in Figs. 2(b), 2(c), 3(b), and 3(c), the calculated quadrupolar susceptibilities are clearly different between the two models at high magnetic fields. Thus, we can distinguish the CEF ground state by checking whether the elastic constants increase or decrease with increasing magnetic field. Here, we present the results of ultrasonic measurements in pulsed magnetic fields with the conclusion that the CEF ground state in SmOs$_4$Sb$_{12}$ is the $\Gamma_{67}$ quartet.

\begin{figure}[t]
\begin{center}
\vspace*{-1mm}
\includegraphics[width=1.0\linewidth]{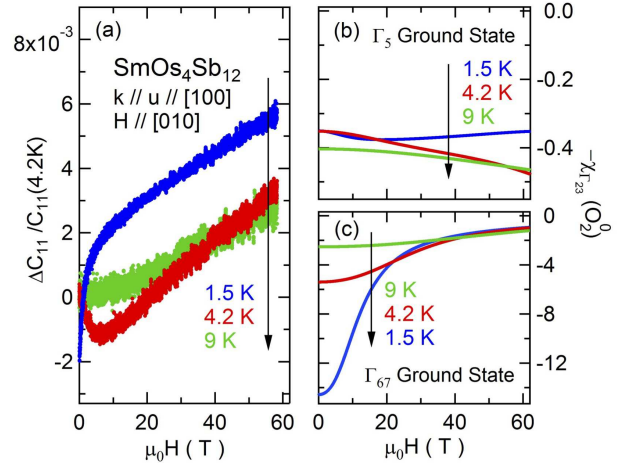}
\end{center}
\caption{(Color online) (a) Relative change in the elastic constant $C_{11}$ versus magnetic field. The zero-field values of the experimental results are calculated on the basis of the temperature dependence of $C_{11}$ reported in Ref. 7.
(b) Magnetic field dependence of the quadrupolar susceptibility ${\chi}_{{\Gamma}_{23}}$($O^{0}_{2}$), calculated for the $\Gamma_{5}$ ground state and the $\Gamma_{67}$ first excited state separated by a gap ${\Delta}$ ${\approx}$ 38 K. (c) The same as in (b) but for the $\Gamma_{67}$ ground state and the $\Gamma_{5}$ first excited state separated by ${\Delta}$ ${\approx}$ 20 K (see text for details).
}
\label{f1}
\end{figure}

A single crystal of SmOs$_4$Sb$_{12}$ was grown by the Sb-flux method. The conventional pulse-echo method was employed for the ultrasonic measurements. Pulsed magnetic field experiments were performed at the Dresden High Magnetic Field Laboratory (Helmholtz-Zentrum Dresden-Rossendorf). The maximum magnetic field was 58.3 T, and the whole pulse duration was $\approx$ 200 ms. The repetition rate of the ultrasonic pulse was set to 55 kHz. A superconducting magnet was used for static field measurements up to 17.5 T. Resonance LiNbO$_3$ transducers were employed for exciting and detecting ultrasonic waves.
The ultrasonic frequencies were 106 MHz for the measurements of $C_{11}$, 70 MHz for the measurements of $C_{44}$ at 1.5 and 4.2 K, and 19 MHz for the measurement of $C_{44}$ at 8.7 K in pulsed magnetic fields.
We have not observed any difference between the results for up and down sweeps of the pulsed magnetic field. We conclude that the heating of the sample due to eddy currents is negligible.

Figure 2(a) shows the pulsed field data for ${\it C}_{44}$ at various temperatures. This elastic constant exhibits a broad minimum at  $\approx$ 4 and  $\approx$ 10 T for 1.5 and 4.2 K, respectively. This characteristic feature is confirmed by experimental results obtained in a static magnetic field and shown as black curves in Fig. 2(a). We will discuss the possible origin of this minimum later. At higher magnetic fields, ${\it C}_{44}$ increases with increasing magnetic field. This feature can be explained by the quadrupolar susceptibility calculated for the $\Gamma_{67}$ quartet ground state as shown in Fig. 2(c).  In this case, the magnetic field splits the Kramers doublets of the ground quartet (Zeeman splitting), and the system's quadrupolar degrees of freedom are suppressed. As a result, a reduction in the absolute value of the quadrupolar susceptibility (${\it i.e}$., an increase in the elastic constants) is expected [Fig. 2(c)]. On the other hand, in the case of the $\Gamma_{5}$ doublet ground state, the quadrupolar susceptibility should increase [Fig. 2(b)]. Here, the Zeeman splitting causes the mixing of the wave functions of the $\Gamma_{5}$ ground state and the excited $\Gamma_{67}$ quartet state, and the 4$f$-electron system gains quadrupolar degrees of freedom. 
In Fig. 3(a), the relative change in the elastic constant ${\it C}_{11}$ is shown. The elastic constant ${\it C}_{11} = {\it C}_B$ + $\frac{4}{3} \frac{({\it C}_{11}-C_{12})}{2}$ includes not only the bulk modulus $C_{\rm B}$, which corresponds to volume change with the $\Gamma_1$ symmetry, but also $\frac{({\it C}_{11}-C_{12})}{2}$, which corresponds to the $\Gamma_{23}$ symmetry, related to the $O_2^0$ quadrupole. The calculated $\Gamma_{23}$-type quadrupolar susceptibility is shown in Figs. 3(b) and 3(c) for the two different ground states. 
Here, we compare $C_{11}$ with the calculated quadrupolar susceptibility ${\chi}_{\Gamma_{23}}$, assuming that the change in the $\Gamma_{23}$ component of the elastic constant, $\frac{({\it C}_{11}-C_{12})}{2}$, is dominant. As can be seen, again our high-field experimental results for $C_{11}$ can be explained well by the ${\Gamma}_{67}$ ground-state model. To estimate the magnetic field dependence of $C_{\rm B}$ more precisely, measurements of the thermal expansion in magnetic fields are necessary.

\begin{figure}[t]
\begin{center}
\includegraphics[width=1.0\linewidth]{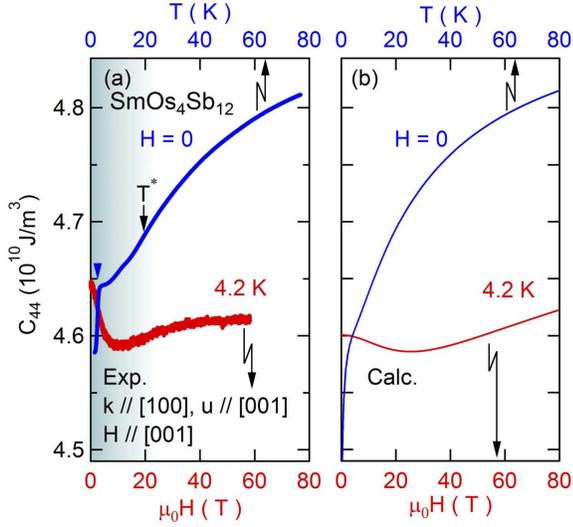}
\end{center}
\caption{(Color online) Red curves show the measured (a) and calculated (b) magnetic field dependences of $C_{44}$ at 4.2 K. Blue curves represent the measured (a) and calculated (b) temperature dependences of $C_{44}$ at 0 T. The calculations are based on the quadrupolar susceptibility for $\Gamma_4$ symmetry and a $\Gamma_{67}$ ground state (see text for details). The colored background region in (a) indicates the temperature and magnetic field regions where 4$f$ electrons are considered to be less localized. 
}
\label{f1}
\end{figure}

Next, we discuss the results obtained at low magnetic fields. Here, a local minimum in ${\it C}_{11}$ at 4.2 K and a relatively large decrease in ${\it C}_{44}$ have been observed [Figs. 2(a) and 3(a)]. To explain these features, we consider two origins. The first one is the magnetoelastic coupling due to an induced weak ferromagnetic moment, which is induced spontaneously below $T_{\textrm C}$ $\approx$ 2.5 K at zero magnetic field \cite{Sanada,Yuhasz}. Since a magnetic field above 10 T polarizes the ferromagnetic moments, we can treat the system as a polarized paramagnetic phase \cite{Sanada}. In this case, we can neglect this magnetoelastic effect and consider only CEF effects above 20 T. 

The other possible origin is the Kondo screening, which is due to $c$--$f$ hybridization and can be suppressed by a magnetic field. In Figs. 4(a) and 4(b), we compare the magnetic field and temperature sweep results to estimate the role of the Kondo effect in this system and to verify the limit of applicability of a localized electron picture. Figure 4(a) shows the magnetic field dependence (bottom axis) and temperature dependence (top axis) of the elastic constant ${\it C}_{44}$. Figure 4(b) shows the results of  the calculation of $C_{44}$, based on the quadrupolar susceptibility for $O_{yz}$ ($=J_yJ_z+J_zJ_y$).

In our analysis, we took into account the temperature-dependent mixed-valence state by multiplying the quadrupolar susceptibility by the ratio of Sm$^{3+}$ to Sm$^{2+}$, $r_v$ = 0.83 -- 0.76. Note that Sm$^{2+}$ ($\vector{J}=0$) has no contribution to the quadrupolar susceptibility. The temperature dependence of the average valence of the Sm ions is assumed to be linear between +2.83 at 150 K and +2.76 at 20 K and constant below 20 K in accordance with Ref. 11. The elastic constant $C_{44}$ = $C_{44}^0$- $\frac{N{g_{{\Gamma}_4}}^2r_v{\chi}_{{\Gamma}_4}}{(1+g^{\prime}_{{\Gamma}_4}r_v{\chi}_{{\Gamma}_4})}$ is calculated with $C_{44}^0 = 4.91{\times}10^{10}$ J/m$^3$, $N = 2.50{\times}10^{27}$ ${\rm m}^{-3}$ \cite{Lattice_Constant}, $|g_{{\Gamma}_4}| = 435$ K, and $g^{\prime}_{{\Gamma}_4} = -1$ K, where $C_{44}^0$, $N$, $g_{{\Gamma}_4}$, $g^{\prime}_{{\Gamma}_4}$, $r_v$, and ${\chi}_{{\Gamma}_4}$ are the background elastic constant, the number of magnetic ions per unit volume, the quadrupole-strain coupling constant, the intersite quadrupole-quadrupole coupling constant, the ratio of Sm$^{3+}$ to Sm$^{2+}$, and the quadrupolar susceptibility, respectively. In the calculation of the magnetic field dependence, we used the average valence at zero magnetic field, since there is no previous report on the magnetic field dependence of the valence state of the Sm ion in SmOs$_4$Sb$_{12}$. Comparison between Figs. 4(a) and 4(b) shows that the quantitative agreement between both temperature sweep and magnetic field sweep is obvious, except at low temperatures and low magnetic fields. Below $T^\ast$ $\approx$ 20 K, the experimental data decrease less markedly towards low temperatures: however, the decrease in the calculated elastic constant continues below 20 K. As discussed in Ref. 7, this deviation can be explained by a modified quadrupolar susceptibility picture that includes a crossover to a Kondo-singlet state, where 4$f$ electrons lose their localized character at low temperatures at zero magnetic field. On the other hand, the amount of change in the experimental results of ${\it C}_{44}$ is also comparable to that of the calculation for $H >$ 20 T and $T >$ 20 K. This means that a significant response of quadrupolar degrees of freedom is observed at high fields, while they recover their localized character. Thus, we conclude that the softening and minima in the elastic constants versus magnetic field at low temperatures can be explained by a combination of the above-mentioned two effects: the recovery of the localized character and the magnetoelastic coupling. To carry out a quantitative analysis, measurements of elastic constants with other geometries, such as $\frac{({\it C}_{11}-C_{12})}{2}$ or $C_{44}$ for other magnetic-field directions, are necessary. 

Finally, we discuss how robust the heavy-fermion state of SmOs$_4$Sb$_{12}$ is in a magnetic field. According to the above discussion, we conclude that a typical Kondo effect of magnetic origin exists in the present compound, which is suppressed by magnetic fields, ${\it i.e}$., the system recovers the localized character of 4$f$ electrons by applying magnetic fields. Indeed, the signature of the Kondo effect has also been found in NQR measurements \cite{NQR_PRL}. 
On the other hand, the robustness of the Sommerfeld coefficient has also been reported, which implies that the present compound possesses a magnetically stable heavy-fermion state \cite{Sanada}. This feature is, however, not verified above 8 T. Our ultrasonic results suggest that magnetic fields above 20 T are sufficient to suppress the Kondo effect. To gain more information on the relationship between the large $\gamma$ and the Kondo-like screened state of this material, more studies, such as magnetization or resistivity measurements in the high magnetic field $\mu_0H > 20$ T are necessary.

In summary, we have performed ultrasonic measurements of SmOs$_4$Sb$_{12}$ in pulsed magnetic fields of up to 58 T. The magnetic field dependences of the elastic constants suggest that the present system possesses a $\Gamma_{67}$ quartet CEF ground state. In our CEF analysis, we assume a simple localized picture of 4$f$ electrons, which qualitatively describes our experimental results obtained above 20 T and at high temperatures as well. Moreover, our results show that the Kondo-like screening due to $c$--$f$ hybridization dominates the low-temperature and low-magnetic-field physics of SmOs$_4$Sb$_{12}$. We conclude that the system recovers the localized character of 4$f$ electrons at magnetic fields above 20 T. \\

\begin{acknowledgments}
We acknowledge the support of the HLD at HZDR, member of the European Magnetic Field Laboratory (EMFL).  Research at HZDR was supported by JSPS KAKENHI Grant No. 26400342 and the Strategic Young Researcher Overseas Visits Program for Accelerating Brain Circulation from the Japan Society for the Promotion of Science. Single-crystal growth and characterization at UCSD were supported by the U. S. Department of Energy, Office of Basic Energy Science, Division of Materials Sciences and Engineering under Grant No. DEFG02-04-ER46105. Research at California State University, Fresno is supported by NSF DMR-1506677.
\end{acknowledgments}

\end{document}